\documentclass[12pt]{article}
\usepackage{epsf}
\def\beq{\begin{equation}}
\def\eeq{\end{equation}}
\def\bea{\begin{eqnarray}}
\def\eea{\end{eqnarray}}

\def\vel{\left|}
\def\ver{\right|}

\def\ga{\left(}
\def\dr{\right)}

\def\rar{\rightarrow}

\def\la{\langle}
\def\ra{\rangle}
\def\ba{\begin{array}}
\def\ea{\end{array}}

\def\ds{\displaystyle}

\title{ {\bf
$B\rightarrow K^* l^+ l^-$ decay in the two Higgs doublet model with flavor 
changing neutral currents}}
\author{\vspace{1cm}\\
         {\bf T. M. Aliev} , \\
         Physics Department, Middle East Technical University \\
         Ankara, Turkey\\        \vspace{5mm}\\
        {\bf E. O. Iltan}
        \thanks{E-mail address:
        eiltan@heraklit.physics.metu.edu.tr}
 \\
        Physics Department, Middle East Technical University \\
        Ankara, Turkey\\}

\date{}

\begin{document}
\setlength{\baselineskip}{24pt}
\maketitle
\setlength{\baselineskip}{7mm}
\begin{abstract}
We study the decay width, forward-backward asymmetry and  
the longitudinal lepton polarization for the exclusive decay 
$B\rightarrow K^* l^+ l^-$ in the two Higgs doublet model with three level
flavor changing neutral currents (model III) and analyse the dependencies of 
these quantities on the the selected parameters, $\xi^{U,D}$, of model III  
including the next to leading  QCD corrections. It is found that to 
look for charged Higgs effects, the measurements of the branching ratio, 
forward-backward asymmetry and  the longitudinal lepton polarization  
for the decay$B\rightarrow K^* l^+ l^-$ are promising.
\end{abstract} 
\thispagestyle{empty}
\newpage
\setcounter{page}{1}

\section{Introduction}
Rare B meson decays induced by flavor changing neutral current (FCNC) 
$b\rightarrow s$ transition represent an important channel to test the 
Standard model (SM) at loop level. They provide a comprehensive information 
in the determination of the fundamental parameters, such as 
Cabbibo-Kobayashi-Maskawa (CKM) matrix elements, leptonic decay constants,
etc. and open a window to investigate the physics beyond the SM, such as 
two Higgs Doublet model (2HDM), Minimal Supersymmetric extension of the SM
(MSSM) \cite{Hewett}, etc. Measurement of the Branching ratios ($Br$)
of the inclusive $B\rightarrow X_s\gamma$ \cite{cleo} and the exclusive
$B\rightarrow K^*\gamma$ \cite{rammar} decays stimulated the study of the 
rare B meson decays in a new force. 

Currently, the main interest is focused on 
the rare decays where the SM predicts large $Br$, which is measurable in 
the near future. $B\rightarrow K^* l^+ l^-$ decay is one of the candidate 
of these decays and it is described by $b\rightarrow s l^+l^-$ transition at 
the quark level. In the literature 
\cite{R4}- \cite{R10} this transition has been investigated extensively in 
the SM, 2HDM and MSSM. The forward-backward asymmetry of dileptons  is 
another quantity which provides information on the short distance 
contributions. Further, the longitudinal lepton polarization includes 
different combinations of the Wilson coefficients $C^{eff}_7$,$C^{eff}_9$ 
and $C_{10}$ and it can shed light on the investigation of their signs in 
the SM and new physics beyond.  
 
The theoretical analysis of exclusive decays is more complicated due to the
hadronic form factors, which brings sustantial uncertainity in the calculations. 
However, it is well known that the experimental investigation of inclusive 
decays is more difficult compared those of exclusive ones. 
Therefore we will consider the exclusive $B\rightarrow K^* l^+ l^-$ in
the present work.

The calculation of the physical observables in the hadronic level needs
non-perturbative methods to determine the matrix elements of the quark 
level effective Hamiltonian between the hadronic states. This problem has 
been studied in the framework of different approaches such as 
relativistic quark model by the light-front formalism \cite{R10}, chiral 
theory \cite{R18}, three point QCD sum rules method \cite{R19},  effective heavy quark theory 
\cite{R21} and light cone QCD sum rules \cite{alsav2,braun}. 

In this work, we present the next to leading order (NLO) QCD corrected 
effective Hamiltonian in the 2HDM with flavor changing neutral currents
(model III) for the inclusive decay $b\rightarrow s l^+ l^-$ and calculate
the differential $Br$  of the exclusive $B\rightarrow K^* l^+ l^-$ decay. 
Further,we study the forward-backward asymmetry ($A_{FB}$) and the 
longitudinal lepton polarization ($P_L$). 

The paper is organized as follows:
In Section 2, we give the NLO QCD corrected Hamiltonian and corresponding 
matrix element for the inclusive $b\rightarrow s\gamma$ decay and calculate 
the matrix element. Section 3 is devoted to the calculation of the 
differential $Br$, $A_{FB}$ and $P_L$ of the exclusive 
$B\rightarrow K^* l^+ l^-$ decay . In Section 4 we analyse the dependencies 
of the $Br$, $A_{FB}$ and $P_L$ on the the Yukawa couplings 
$\bar{\xi}_{bb}^{D}$, $\bar{\xi}_{tt}^{U}$ and estimate the 
$Br(B\rightarrow K^* l^+ l^-)$ for the model III.

\section{\bf QCD corrected short-distance contributions in 
the model III for the decay $b\rightarrow s l^+ l^-$ with additional 
long-distance effects }
In this section we present the NLO QCD corrections to the 
inclusive $b\rightarrow s l^+ l^- $ ($l=e,\mu$)
decay amplitude in the 2HDM with tree level neutral currents (model III). 
Before presenting the details of calculations, we would like to give the
essential points of this model.

The Yukawa interaction in this general case is
\begin{eqnarray}
{\cal{L}}_{Y}=\eta^{U}_{ij} \bar{Q}_{i L} \tilde{\phi_{1}} U_{j R}+
\eta^{D}_{ij} \bar{Q}_{i L} \phi_{1} D_{j R}+
\xi^{U}_{ij} \bar{Q}_{i L} \tilde{\phi_{2}} U_{j R}+
\xi^{D}_{ij} \bar{Q}_{i L} \phi_{2} D_{j R} + h.c. \,\,\, ,
\label{lagrangian}
\end{eqnarray}
where $L$ and $R$ denote chiral projections $L(R)=1/2(1\mp \gamma_5)$,
$\phi_{i}$ for $i=1,2$, are the two scalar doublets, $\eta^{U,D}_{ij}$
and $\xi^{U,D}_{ij}$ are the matrices of the Yukawa couplings.
The Flavor Changing (FC) part of the interaction can be written as 
\begin{eqnarray}
{\cal{L}}_{Y,FC}=
\xi^{U}_{ij} \bar{Q}_{i L} \tilde{\phi_{2}} U_{j R}+
\xi^{D}_{ij} \bar{Q}_{i L} \phi_{2} D_{j R} + h.c. \,\, ,
\label{lagrangianFC}
\end{eqnarray}
with the choice of $\phi_1$ and $\phi_2$
\begin{eqnarray}
\phi_{1}=\frac{1}{\sqrt{2}}\left[\left(\begin{array}{c c} 
0\\v+H^{0}\end{array}\right)\; + \left(\begin{array}{c c} 
\sqrt{2} \chi^{+}\\ i \chi^{0}\end{array}\right) \right]\, ; 
\phi_{2}=\frac{1}{\sqrt{2}}\left(\begin{array}{c c} 
\sqrt{2} H^{+}\\ H_1+i H_2 \end{array}\right) \,\, .
\label{choice}
\end{eqnarray}
Here the vacuum expectation values are,  
\begin{eqnarray}
<\phi_{1}>=\frac{1}{\sqrt{2}}\left(\begin{array}{c c} 
0\\v\end{array}\right) \,  \, ; 
<\phi_{2}>=0 \,\, ,
\label{choice2}
\end{eqnarray}
and the couplings  $\xi^{U,D}$ for the FC charged interactions are
\begin{eqnarray}
\xi^{U}_{ch}&=& \xi_{neutral} \,\, V_{CKM} \nonumber \,\, ,\\
\xi^{D}_{ch}&=& V_{CKM} \,\, \xi_{neutral} \,\, ,
\label{ksi1} 
\end{eqnarray}
where  $\xi^{U,D}_{neutral}$ 
\footnote{In all next discussion we denote $\xi^{U,D}_{neutral}$ 
as $\xi^{U,D}_{N}$.} 
is defined by the expression
\begin{eqnarray}
\xi^{U,D}_{N}=(V_L^{U,D})^{-1} \xi^{U,D} V_R^{U,D}\,\, .
\label{ksineut}
\end{eqnarray}
Note that the charged couplings appear as a linear combinations of neutral 
couplings multiplied by $V_{CKM}$ matrix elements (more details see
\cite{soni}). 

The next step is the calculation of the QCD corrections to
$b\rightarrow  s l^+ l^-$ decay amplitude in 
the model III. 
The appropriate framework is that of an effective theory obtained by 
integrating out the heavy degrees of freedom.
In the present case, $t$ quark, $W^{\pm}, H^{\pm}, H_{1}$, and $H_{2}$ 
bosons are the heavy degrees of freedom. Here $H^{\pm}$ denote charged, 
$H_{1}$ and $H_{2}$ denote neutral Higgs bosons. The QCD corrections 
are done through matching the full theory with the effective low energy 
theory at the high scale $\mu=m_{W}$ and evaluating the Wilson coefficients 
from $m_{W}$ down to the lower scale $\mu\sim O(m_{b})$.
Note that we choose the higher scale as $\mu=m_{W}$ since the evaluation 
from the scale $\mu=m_{H^{\pm}}$ to $\mu=m_{W}$ gives negligible
contribution to the Wilson coefficients since the charged Higgs boson is 
heavy enough from the current theoretical restrictions, for example  
$m_{H^{\pm}} \geq 340\, GeV$ \cite{ciuchini2}, 
$m_{H^{\pm}} \geq 480\, GeV$ \cite{gudalil}.
      
The effective Hamiltonian relevant for our process is
\begin{eqnarray}
{\cal{H}}_{eff}=-4 \frac{G_{F}}{\sqrt{2}} V_{tb} V^{*}_{ts} 
\sum_{i}C_{i}(\mu) O_{i}(\mu) \, \, ,
\label{hamilton}
\end{eqnarray}
where the $O_{i}$ are operators given in eqs.~(\ref{op1}),~(\ref{op2}) 
and the $C_{i}(\mu)$ are Wilson coefficients
renormalized at the scale $\mu$. The coefficients $C_{i}(\mu)$ are calculated 
perturbatively.

The operator basis is the extension of the 
one used for the SM and 2HDM (model I and II) \cite{Grinstein2, misiak}
with the additional flipped chirality partners, similar to the ones used
for the $b\rightarrow s \gamma$ decay in the  model III \cite{alil} and 
$SU(2)_L\times SU(2)_R\times U(1)$ extensions of the SM \cite{cho}:
\begin{eqnarray}
 O_1 &=& (\bar{s}_{L \alpha} \gamma_\mu c_{L \beta})
               (\bar{c}_{L \beta} \gamma^\mu b_{L \alpha}), \nonumber   \\
 O_2 &=& (\bar{s}_{L \alpha} \gamma_\mu c_{L \alpha})
               (\bar{c}_{L \beta} \gamma^\mu b_{L \beta}),  \nonumber   \\
 O_3 &=& (\bar{s}_{L \alpha} \gamma_\mu b_{L \alpha})
               \sum_{q=u,d,s,c,b}
               (\bar{q}_{L \beta} \gamma^\mu q_{L \beta}),  \nonumber   \\
 O_4 &=& (\bar{s}_{L \alpha} \gamma_\mu b_{L \beta})
                \sum_{q=u,d,s,c,b}
               (\bar{q}_{L \beta} \gamma^\mu q_{L \alpha}),   \nonumber  \\
 O_5 &=& (\bar{s}_{L \alpha} \gamma_\mu b_{L \alpha})
               \sum_{q=u,d,s,c,b}
               (\bar{q}_{R \beta} \gamma^\mu q_{R \beta}),   \nonumber  \\
 O_6 &=& (\bar{s}_{L \alpha} \gamma_\mu b_{L \beta})
                \sum_{q=u,d,s,c,b}
               (\bar{q}_{R \beta} \gamma^\mu q_{R \alpha}),  \nonumber   \\  
 O_7 &=& \frac{e}{16 \pi^2}
          \bar{s}_{\alpha} \sigma_{\mu \nu} (m_b R + m_s L) b_{\alpha}
                {\cal{F}}^{\mu \nu},                             \nonumber  \\
 O_8 &=& \frac{g}{16 \pi^2}
    \bar{s}_{\alpha} T_{\alpha \beta}^a \sigma_{\mu \nu} (m_b R + m_s L)  
          b_{\beta} {\cal{G}}^{a \mu \nu} \nonumber \,\, , \\  
 O_9 &=& \frac{e}{16 \pi^2}
          (\bar{s}_{L \alpha} \gamma_\mu b_{L \alpha})
              (\bar{l} \gamma_\mu l)  \,\, ,    \nonumber    \\
 O_{10} &=& \frac{e}{16 \pi^2}
          (\bar{s}_{L \alpha} \gamma_\mu b_{L \alpha})
              (\bar{l} \gamma_\mu \gamma_{5} l)  \,\, ,    \nonumber  \\
 O_{11} &=& (\bar{s}_{L \alpha} \gamma_\mu c_{L \beta})
               (\bar{c}_{R \beta} \gamma^\mu b_{R \alpha}), \nonumber   \\
 O_{12} &=& (\bar{s}_{L \alpha} \gamma_\mu c_{L \alpha})
(\bar{c}_{R \beta} \gamma^\mu b_{R \beta}),
\label{op1}
\end{eqnarray}
and the second operator set $O'_{1} - O'_{12}$ which are 
flipped chirality partners of $O_{1} - O_{12}$:
\begin{eqnarray}
 O'_1 &=& (\bar{s}_{R \alpha} \gamma_\mu c_{R \beta})
               (\bar{c}_{R \beta} \gamma^\mu b_{R \alpha}), \nonumber   \\
 O'_2 &=& (\bar{s}_{R \alpha} \gamma_\mu c_{R \alpha})
               (\bar{c}_{R \beta} \gamma^\mu b_{R \beta}),  \nonumber   \\
 O'_3 &=& (\bar{s}_{R \alpha} \gamma_\mu b_{R \alpha})
               \sum_{q=u,d,s,c,b}
               (\bar{q}_{R \beta} \gamma^\mu q_{R \beta}),  \nonumber   \\
 O'_4 &=& (\bar{s}_{R \alpha} \gamma_\mu b_{R \beta})
                \sum_{q=u,d,s,c,b}
               (\bar{q}_{R \beta} \gamma^\mu q_{R \alpha}),   \nonumber  \\
 O'_5 &=& (\bar{s}_{R \alpha} \gamma_\mu b_{R \alpha})
               \sum_{q=u,d,s,c,b}
               (\bar{q}_{L \beta} \gamma^\mu q_{L \beta}),   \nonumber  \\
 O'_6 &=& (\bar{s}_{R \alpha} \gamma_\mu b_{R \beta})
                \sum_{q=u,d,s,c,b}
               (\bar{q}_{L \beta} \gamma^\mu q_{L \alpha}),  \nonumber   \\  
 O'_7 &=& \frac{e}{16 \pi^2}
          \bar{s}_{\alpha} \sigma_{\mu \nu} (m_b L + m_s R) b_{\alpha}
                {\cal{F}}^{\mu \nu},                             \nonumber       \\
 O'_8 &=& \frac{g}{16 \pi^2}
    \bar{s}_{\alpha} T_{\alpha \beta}^a \sigma_{\mu \nu} (m_b L + m_s R)  
          b_{\beta} {\cal{G}}^{a \mu \nu}, \nonumber \\ 
 O'_9 &=& \frac{e}{16 \pi^2}
          (\bar{s}_{R \alpha} \gamma_\mu b_{R \alpha})
              (\bar{l} \gamma_\mu l)  \,\, ,    \nonumber  \\
 O'_{10} &=& \frac{e}{16 \pi^2}
          (\bar{s}_{R \alpha} \gamma_\mu b_{R \alpha})
              (\bar{l} \gamma_\mu \gamma_{5} l)  \,\, ,    \nonumber \\
 O'_{11} &=& (\bar{s}_{R \alpha} \gamma_\mu c_{R \beta})
               (\bar{c}_{L \beta} \gamma^\mu b_{L \alpha})\,\, , \nonumber   \\
 O'_{12} &=& (\bar{s}_{R \alpha} \gamma_\mu c_{R \alpha})
(\bar{c}_{L \beta} \gamma^\mu b_{L \beta})\,\, ,
\label{op2}
\end{eqnarray}
where  
$\alpha$ and $\beta$ are $SU(3)$ colour indices and
${\cal{F}}^{\mu \nu}$ and ${\cal{G}}^{\mu \nu}$
are the field strength tensors of the electromagnetic and strong
interactions, respectively.

Denoting the Wilson coefficients for the SM with $C_{i}^{SM}(m_{W})$ and the
additional charged Higgs contribution with $C_{i}^{H}(m_{W})$, 
we have the initial values for the first set of operators 
(eq.(\ref{op1})) \cite{Grinstein1,alil} 
\begin{eqnarray}
C^{SM}_{1,3,\dots 6,11,12}(m_W)&=&0 \nonumber \, \, , \\
C^{SM}_2(m_W)&=&1 \nonumber \, \, , \\
C_7^{SM}(m_W)&=&\frac{3 x^3-2 x^2}{4(x-1)^4} \ln x+
\frac{-8x^3-5 x^2+7 x}{24 (x-1)^3} \nonumber \, \, , \\
C_8^{SM}(m_W)&=&-\frac{3 x^2}{4(x-1)^4} \ln x+
\frac{-x^3+5 x^2+2 x}{8 (x-1)^3}\nonumber \, \, , \\ 
C_9^{SM}(m_W)&=&-\frac{1}{sin^2\theta_{W}} B(x) +
\frac{1-4 \sin^2 \theta_W}{\sin^2 \theta_W} C(x)-D(x)+\frac{4}{9}, \nonumber \, \, , \\
C_{10}^{SM}(m_W)&=&\frac{1}{\sin^2\theta_W}
(B(x)-C(x))\nonumber \,\, , \\
C^{H}_{1,\dots 6,11,12}(m_W)&=&0 \nonumber \, \, , \\
C_7^{H}(m_W)&=&\frac{1}{m_{t}^2} \,
(\bar{\xi}^{U}_{N,tt}+\bar{\xi}^{U}_{N,tc}
\frac{V_{cs}^{*}}{V_{ts}^{*}}) \, (\bar{\xi}^{U}_{N,tt}+\bar{\xi}^{U}_{N,tc}
\frac{V_{cb}}{V_{tb}}) F_{1}(y)\nonumber  \, \, , \\
&+&\frac{1}{m_t m_b} \, (\bar{\xi}^{U}_{N,tt}+\bar{\xi}^{U}_{N,tc}
\frac{V_{cs}^{*}}{V_{ts}^{*}}) \, (\bar{\xi}^{D}_{N,bb}+\bar{\xi}^{D}_{N,sb}
\frac{V_{ts}}{V_{tb}}) F_{2}(y)
\nonumber  \, \, , \\
C_8^{H}(m_W)&=&\frac{1}{m_{t}^2} \,
(\bar{\xi}^{U}_{N,tt}+\bar{\xi}^{U}_{N,tc}
\frac{V_{cs}^{*}}{V_{ts}^{*}}) \, (\bar{\xi}^{U}_{N,tt}+\bar{\xi}^{U}_{N,tc}
\frac{V_{cb}}{V_{tb}})G_{1}(y)
\nonumber  \, \, , \\
&+&\frac{1}{m_t m_b} \, (\bar{\xi}^{U}_{N,tt}+\bar{\xi}^{U}_{N,tc}
\frac{V_{cs}^{*}}{V_{ts}^{*}}) \, (\bar{\xi}^{D}_{N,bb}+\bar{\xi}^{U}_{N,sb}
\frac{V_{ts}}{V_{tb}}) G_{2}(y) \nonumber\, \, , \\
C_9^{H}(m_W)&=&\frac{1}{m_{t}^2} \,
(\bar{\xi}^{U}_{N,tt}+\bar{\xi}^{U}_{N,tc}
\frac{V_{cs}^{*}}{V_{ts}^{*}}) \, (\bar{\xi}^{U}_{N,tt}+\bar{\xi}^{U}_{N,tc}
\frac{V_{cb}}{V_{tb}}) H_{1}(y)
\nonumber  \, \, , \\
C_{10}^{H}(m_W)&=&\frac{1}{m_{t}^2} \,
(\bar{\xi}^{U}_{N,tt}+\bar{\xi}^{U}_{N,tc}
\frac{V_{cs}^{*}}{V_{ts}^{*}}) \, (\bar{\xi}^{U}_{N,tt}+\bar{\xi}^{U}_{N,tc}
\frac{V_{cb}}{V_{tb}}) L_{1}(y) \, \, , 
\label{CoeffH}
\end{eqnarray}
and for the second set of operators eq.~(\ref{op2}), 
\begin{eqnarray}
C^{\prime SM}_{1,\dots 12}(m_W)&=&0 \nonumber \, \, , \\
C^{\prime H}_{1,\dots 6,11,12}(m_W)&=&0 \nonumber \, \, , \\
C^{\prime H}_7(m_W)&=&\frac{1}{m_t^2} \,
(\bar{\xi}^{D}_{N,bs}\frac{V_{tb}}{V_{ts}^{*}}+\bar{\xi}^{D}_{N,ss})
\, (\bar{\xi}^{D}_{N,bb}+\bar{\xi}^{D}_{N,sb}
\frac{V_{ts}}{V_{tb}}) F_{1}(y)
\nonumber  \, \, , \\
&+& \frac{1}{m_t m_b}\, (\bar{\xi}^{D}_{N,bs}\frac{V_{tb}}{V_{ts}^{*}}
+\bar{\xi}^{D}_{N,ss}) \, (\bar{\xi}^{U}_{N,tt}+\bar{\xi}^{U}_{N,tc}
\frac{V_{cb}}{V_{tb}}) F_{2}(y)
\nonumber  \, \, , \\
C^{\prime H}_8 (m_W)&=&\frac{1}{m_t^2} \,
(\bar{\xi}^{D}_{N,bs}\frac{V_{tb}}{V_{ts}^{*}}+\bar{\xi}^{D}_{N,ss})
\, (\bar{\xi}^{D}_{N,bb}+\bar{\xi}^{D}_{N,sb}
\frac{V_{ts}}{V_{tb}}) G_{1}(y)
\nonumber  \, \, , \\
&+&\frac{1}{m_t m_b} \, (\bar{\xi}^{D}_{N,bs}\frac{V_{tb}}{V_{ts}^{*}}
+\bar{\xi}^{D}_{N,ss}) \, (\bar{\xi}^{U}_{N,tt}+\bar{\xi}^{U}_{N,tc}
\frac{V_{cb}}{V_{tb}}) G_{2}(y)
\nonumber \,\, ,\\
C^{\prime H}_9(m_W)&=&\frac{1}{m_t^2} \,
(\bar{\xi}^{D}_{N,bs}\frac{V_{tb}}{V_{ts}^{*}}+\bar{\xi}^{D}_{N,ss})
\, (\bar{\xi}^{D}_{N,bb}+\bar{\xi}^{D}_{N,sb}
\frac{V_{ts}}{V_{tb}}) H_{1}(y)
\nonumber  \, \, , \\
C^{\prime H}_{10} (m_W)&=&\frac{1}{m_t^2} \,
(\bar{\xi}^{D}_{N,bs}\frac{V_{tb}}{V_{ts}^{*}}+\bar{\xi}^{D}_{N,ss})
\, (\bar{\xi}^{D}_{N,bb}+\bar{\xi}^{D}_{N,sb}
\frac{V_{ts}}{V_{tb}}) L_{1}(y)
\,\, ,
\label{CoeffH2}
\end{eqnarray}
where $x=m_t^2/m_W^2$ and $y=m_t^2/m_{H^{\pm}}^2$.
In eqs.~(\ref{CoeffH}) and (\ref{CoeffH2}) we used the redefinition
\begin{eqnarray}
\xi^{U,D}=\sqrt{\frac{4 G_{F}}{\sqrt{2}}} \,\, \bar{\xi}^{U,D}\,\, .
\label{ksidefn}
\end{eqnarray}  
The functions $B(x)$, $C(x)$, $D(x)$, $F_{1(2)}(y)$, $G_{1(2)}(y)$, 
$H_{1}(y)$ and $L_{1}(y)$ are given as
\begin{eqnarray}
B(x)&=&\frac{1}{4}\left[\frac{-x}{x-1}+\frac{x}{(x-1)^2} \ln
x\right] \nonumber \,\, , \\
C(x)&=&\frac{x}{4}\left[\frac{x/2-3}{x-1}+\frac{3x/2+1}{(x-1)^2}
       \ln x \right] \nonumber \,\, , \\
D(x)&=&\frac{-19x^3/36+25x^2/36}{(x-1)^3}+
       \frac{-x^4/6+5x^3/3-3x^2+16x/9-4/9}{(x-1)^4}\ln x 
\nonumber \,\, ,\\
F_{1}(y)&=& \frac{y(7-5y-8y^2)}{72 (y-1)^3}+\frac{y^2 (3y-2)}{12(y-1)^4}
\,\ln y \nonumber  \,\, , \\ 
F_{2}(y)&=& \frac{y(5y-3)}{12 (y-1)^2}+\frac{y(-3y+2)}{6(y-1)^3}\, \ln y 
\nonumber  \,\, ,\\ 
G_{1}(y)&=& \frac{y(-y^2+5y+2)}{24 (y-1)^3}+\frac{-y^2} {4(y-1)^4} \, \ln y
\nonumber  \,\, ,\\ 
G_{2}(y)&=& \frac{y(y-3)}{4 (y-1)^2}+\frac{y} {2(y-1)^3} \, \ln y 
\nonumber\,\, ,\\
H_{1}(y)&=& \frac{1-4 sin^2\theta_W}{sin^2\theta_W}\,\, \frac{x
y}{8}\,\left[ 
\frac{1}{y-1}-\frac{1}{(y-1)^2} \ln y \right]-y \left[\frac{47 y^2-79 y+38}{108
(y-1)^3}-\frac{3 y^3-6 y+4}{18(y-1)^4} \ln y \right] 
\nonumber  \,\, , \\ 
L_{1}(y)&=& \frac{1}{sin^2\theta_W} \,\,\frac{x y}{8}\, \left[-\frac{1}{y-1}+
\frac{1}{(y-1)^2} \ln y \right]
\nonumber  \,\, .\\ 
\label{F1G1}
\end{eqnarray}
Note that we neglect the contributions due to the 
neutral Higgs bosons since their interactions include Yukawa
couplings which should be very small (see the discussion section for details).

For the initial values of the Wilson coefficients in the model III  
(eqs. (\ref{CoeffH})and (\ref{CoeffH2})), we have 
\begin{eqnarray}
C^{2HDM}_{1,3,\dots 6,11,12}(m_W)&=&0 \nonumber \, \, , \\
C_2^{2HDM}(m_W)&=&1 \nonumber \, \, , \\
C_7^{2HDM}(m_W)&=&C_7^{SM}(m_W)+C_7^{H}(m_W) \nonumber \, \, , \\
C_8^{2HDM}(m_W)&=&C_8^{SM}(m_W)+C_8^{H}(m_W) \nonumber \, \, , \\ 
C_9^{2HDM}(m_W)&=&C_9^{SM}(m_W)+C_9^{H}(m_W) \nonumber \, \, , \\
C_{10}^{2HDM}(m_W)&=&C_{10}^{SM}(m_W)+C_{10}^{H}(m_W) \nonumber \, \, , \\ 
\nonumber \\ 
C^{\prime 2HDM}_{1,2,3,\dots 6,11,12}(m_W)&=&0 \nonumber \, \, , \\
C_7^{\prime 2HDM}(m_W)&=&C_7^{\prime SM}(m_W)+C_7^{\prime H}(m_W) \nonumber \, \, , \\
C_8^{\prime 2HDM}(m_W)&=&C_8^{\prime SM}(m_W)+C_8^{\prime H}(m_W) \nonumber \, \, , \\ 
C_9^{\prime 2HDM}(m_W)&=&C_9^{\prime SM}(m_W)+C_9^{\prime H}(m_W) \nonumber \, \, , \\
C_{10}^{\prime 2HDM}(m_W)&=&C_{10}^{\prime SM}(m_W)+C_{10}^{\prime H}(m_W) \, \, . 
\label{Coef2HDM}
\end{eqnarray}

Using the initial values, we can calculate the coefficients 
$C_{i}^{2HDM}$ and $C^{\prime 2HDM}_{i}$ at any lower 
scale with five quark effective theory where large logarithims can be 
summed using the renormalization group and their evaluations are similar 
to the SM case.

The operators 
$O_5$, $O_6$, $O_{11}$ and $O_{12}$ ( $O'_5$, $O'_6$, $O'_{11}$ and $O'_{12}$) give a contribution 
to the leading order matrix element of $b\rightarrow s\gamma$ and  
the magnetic moment type coefficient $C_7^{eff}(\mu)$ ($C_7^{\prime eff}(\mu)$) 
is redefined in the NDR scheme as \cite{alil}:
\begin{eqnarray}
C_{7}^{eff}(\mu)&=&C_{7}^{2HDM}(\mu)+ Q_d \, 
(C_{5}^{2HDM}(\mu) + N_c \, C_{6}^{2HDM}(\mu))\nonumber \, \, , \\
&+& Q_u\, (\frac{m_c}{m_b}\, C_{12}^{2HDM}(\mu) + N_c \, 
\frac{m_c}{m_b}\,C_{11}^{2HDM}(\mu))\nonumber \, \, , \\
C^{\prime eff}_7(\mu)&=& C^{\prime 2HDM}_7(\mu)+Q_{d}\, 
(C^{\prime 2HDM}_5(\mu) + N_c \, C^{\prime 2HDM}_6(\mu))\nonumber \\
&+& Q_u (\frac{m_c}{m_b}\, C_{12}^{\prime 2HDM}(\mu) + N_c \, 
\frac{m_c}{m_b}\,C_{11}^{\prime 2HDM}(\mu))\, \, .
\label{C7eff}
\end{eqnarray}

The NLO corrected coefficients $C_{7}^{2HDM}(\mu)$ and
$C^{\prime 2HDM}_7(\mu)$  are given as 
\begin{eqnarray}
C_{7}^{2HDM}(\mu)&=&C_{7}^{LO, 2HDM}(\mu) 
+\frac{\alpha_s (\mu)}{4\pi} C_7^{(1)\, 2HDM}(\mu) \nonumber \,\, , \\
C^{\prime 2HDM}_7(\mu)&=& C_{7}^{\prime LO, 2HDM}(\mu) + 
\frac{\alpha_s (\mu)}{4\pi} C_7^{\prime (1)\, 2HDM}(\mu)\,\, .
\label{renwils}
\end{eqnarray}

The functions $C_{7}^{LO, 2HDM}(\mu)$ and $C_{7}^{\prime LO, 2HDM}(\mu$)
are the leading order QCD corrected Wilson coefficients
\cite{buras,Grinstein2,misiak}:
\begin{eqnarray} 
C_{7}^{LO, 2HDM}(\mu)&=& \eta^{16/23} C_{7}^{2HDM}(m_{W})+(8/3) 
(\eta^{14/23}-\eta^{16/23}) C_{8}^{2HDM}(m_{W})\nonumber \,\, \\
&+& C_{2}^{2HDM}(m_{W}) \sum_{i=1}^{8} h_{i} \eta^{a_{i}} \,\, , \nonumber
\\
C_{7}^{\prime LO, 2HDM}(\mu)&=& \eta^{16/23} C^{\prime 2HDM}_7(m_{W})+
(8/3) (\eta^{14/23}-\eta^{16/23}) C^{\prime 2HDM}_8(m_{W}) \,\,
\label{LOwils}
\end{eqnarray}
where $\eta =\alpha_{s}(m_{W})/\alpha_{s}(\mu)$, $h_{i}$ and $a_{i}$ are 
the numbers which appear during the evaluation \cite{buras}. 
$C_7^{(1)\, 2HDM}(\mu)$ is the $\alpha_s$ correction to the leading
order result that its explicit form can be found in \cite{ciuchini2,greub2}.
$C_7^{\prime (1)\, 2HDM}(\mu)$ can be obtained by replacing the Wilson 
coefficients in $C_7^{(1)\, 2HDM}(\mu)$ with their primed counterparts. 

Since $O_2$ ($O'_2$) produce dilepton via virtual photon, its Wilson 
coefficient $C_2(\mu)$ ($C^{\prime}_2(\mu)$) and the coefficients 
$C_1(\mu)$, $C_3(\mu)$, ...., $C_6(\mu)$ ($C^{\prime}_2(\mu)$, $C^{\prime}_3
(\mu)$, ..., $C^{\prime}_6(\mu) $) induced by the operator mixing, give 
contributions to $C_9^{eff}(\mu)$ ($C^{\prime eff}_9(\mu)$).
The perturbative part of $C_9^{eff}(\mu)$ \cite{buras,misiak} and 
$C^{\prime eff}_9(\mu)$ including NLO QCD corrections 
are defined in the NDR scheme as:  
\begin{eqnarray} 
C_9^{pert}(\mu)&=& C_9^{2HDM}(\mu) \tilde\eta (\hat s) \nonumber 
\\ &+& h(z, \hat s) \left( 3 C_1(\mu) + C_2(\mu) + 3 C_3(\mu) + 
C_4(\mu) + 3 C_5(\mu) + C_6(\mu) \right) \nonumber \\
&- & \frac{1}{2} h(1, \hat s) \left( 4 C_3(\mu) + 4 C_4(\mu) + 3
C_5(\mu) + C_6(\mu) \right) \\
&- &  \frac{1}{2} h(0, \hat s) \left( C_3(\mu) + 3 C_4(\mu) \right) +
\frac{2}{9} \left( 3 C_3(\mu) + C_4(\mu) + 3 C_5(\mu) + C_6(\mu)
\right) \nonumber \,\, ,
\label{C9eff}
\end{eqnarray}
and
\newpage 
\begin{eqnarray} 
C_9^{\prime\, pert}(\mu)&=& C_9^{\prime 2HDM}(\mu) \tilde\eta(\hat s)\nonumber 
\\ &+& h(z, \hat s) \left( 3 C'_1(\mu) + C'_2(\mu) + 3 C'_3(\mu) + 
C'_4(\mu) + 3 C'_5(\mu) + C'_6(\mu) \right) \nonumber \\
&-& \frac{1}{2} h(1, \hat s) \left( 4 C'_3(\mu) + 4 C'_4(\mu) + 3
C'_5(\mu) + C'_6(\mu) \right) \\
&-&  \frac{1}{2} h(0, \hat s) \left( C'_3(\mu) + 3 C'_4(\mu) \right) +
\frac{2}{9} \left( 3 C'_3(\mu) + C'_4(\mu) + 3 C'_5(\mu) + C'_6(\mu)
\right) \nonumber \, .
\label{C9effp} 
\end{eqnarray}
where $z=\frac{m_c}{m_b}$ a-nd $\hat s=\frac{q^2}{m_b^2}$.
In the above expression $\tilde\eta(\hat s)$ represents the one gluon
correction to the matrix element $O_9$ with $m_s=0$ \cite{misiak} and
the function $h(z,\hat s)$ arises from the one loop contributions of the
four quark operators $O_1, ... ,O_6$ ($O'_1, ... ,O'_6$).  Their explicit
forms can be found in \cite{buras, misiak}.
In addition to the short distance part there exist the long
distance (LD) effects due to the real $\bar{c}c$ in the intermediate states,
i.e. the cascade process $B\rightarrow K^* \psi_i \rightarrow K^* l^+ l^-$
where $i=1,..,6$. In the current literature, there are four different
approaches to take these intermediate states into account:
HQET approach \cite{int1}, AMM approach \cite{ali}, 
LSW approach \cite{int3} and  KS approach \cite{int4}.  
In our calculations, we follow the approach AMM. (We also use 
KS approach in our numerical analysis and see that results obtained in both
methods are closed each other.) 
In this method the resonance $c\bar{c}$ contribution is parametrized 
using a Breit-Wigner shape with normalizations fixed by data, and this
contribution is added to the perturbative one coming from the $c\bar{c}$
loop: 
\begin{eqnarray}
C_9^{eff}(\mu)=C_9^{pert}(\mu)+ Y_{reson}(\hat{s})\,\, ,
\label{C9efftot}
\end{eqnarray}
where $Y_{reson}(\hat{s})$ in NDR scheme is defined as
\begin{eqnarray}
Y_{reson}(\hat{s})&=&-\frac{3}{\alpha^2_{em}}\kappa \sum_{V_i=\psi_i}
\frac{\pi \Gamma(V_i\rightarrow ll)m_{V_i}}{q^2-m_{V_i}+i m_{V_i}
\Gamma_{V_i}} \nonumber \\
& & \left( 3 C_1(\mu) + C_2(\mu) + 3 C_3(\mu) + 
C_4(\mu) + 3 C_5(\mu) + C_6(\mu) \right).
\label{Yres}
\end{eqnarray}
For the expression $C_9^{\prime \, eff}(\mu)$, it is enough to replace
all unprimed coefficients with primed ones.
In eqs. (18) and (19) the phenomenological parameter 
$\kappa=2.3$ is chosen to be able to reproduce the correct value of the 
branching ratio  $Br(B\rightarrow J/\psi X\rightarrow X l\bar{l})=  
Br(B\rightarrow J/\psi X)\, Br(J/\psi\rightarrow X l\bar{l})$ \cite{ali}.
The NLO corrected coefficients $C_i\,\,, i=1,...,6$ can be found in 
\cite{ciuchini2,greub2}. In our numerical calculations we neglect the 
coefficients $C_{5}^{2HDM}(\mu)$, $C_{6}^{2HDM}(\mu)$ and  
$C^{\prime 2HDM}_5(\mu)$, $C^{\prime 2HDM}_6(\mu)$ since they are numerically 
small at $m_b$ scale.

Note that the expressions for unprimed Wilson coefficients in our case 
can be obtained from the results in \cite{greub2} by the following 
replacements:
\begin{eqnarray}
|Y|^2 &\rightarrow& \frac{1}{m_{t}^2} \,
(\bar{\xi}^{U}_{N,tt}+\bar{\xi}^{U}_{N,tc}
\frac{V_{cs}^{*}}{V_{ts}^{*}}) \, (\bar{\xi}^{U}_{N,tt}+\bar{\xi}^{U}_{N,tc}
\frac{V_{cb}}{V_{tb}}) \nonumber \,\, , \\
XY &\rightarrow& \frac{1}{m_t m_b} \, (\bar{\xi}^{U}_{N,tt}+\bar{\xi}^{U}_{N,tc}
\frac{V_{cs}^{*}}{V_{ts}^{*}}) \, (\bar{\xi}^{D}_{N,bb}+\bar{\xi}^{D}_{N,sb}
\frac{V_{ts}}{V_{tb}})
\label{repl}
\end{eqnarray}

To obtain primed coefficients, it is enough to use the primed 
ones at $m_W$ level (eq. (\ref{CoeffH2})) since the evaluation 
of $C_{i}^{\prime}(\mu)$ from $\mu=m_W$ to $\mu=m_b$ is the same as
that of $C_{i}(\mu)$.

For model II (model I) $Y$ and $XY$ are 
\begin{eqnarray}
Y&=&1/tan\beta \,\, (1/tan\beta) \nonumber \,\,, \\
XY&=&1\,\, (-1/tan^2\beta) \,\, .
\end{eqnarray}

At this stage we would like to make the following remark. In \cite{kagan}
QED corrections to the Wilson coefficient $C_7^{eff}$ for 
$B\rightarrow X_s\gamma$ decay are calculated in addition to the NLO
QCD ones and it was shown that QED corrections were almost one order
smaller than NLO QCD ones. We expect that the similar situation 
exists for $C_9^{eff}$. Therefore, in our calculations we neglect them.

Finally, neglecting the strange quark mass, the matrix element for 
$b \rightarrow s \ell^+\ell^-$ decay is obtained as:
\begin{eqnarray}
{\cal M}&=& - \frac{G_F \alpha_{em}}{2\sqrt 2 \pi} V_{tb} V^*_{ts} 
\Bigg\{ \left( \, C_9^{eff}(\mu)\,
\bar s \gamma_\mu (1- \gamma_5) b + 
C_9^{\prime eff}(\mu)\, \bar s \gamma_\mu (1+ \gamma_5) b \, \right)
\,\, \bar \ell \gamma^\mu \ell \nonumber \\
&+& \left( \, C_{10}(\mu) \, \bar s \gamma_\mu (1- \gamma_5) b+
C'_{10}(\mu)\, \bar s \gamma_\mu (1+ \gamma_5) b \, \right) \,\,
\bar \ell \gamma^\mu \gamma_5 \ell   \\
&-& 2 \left( \, C^{eff}_7(\mu)\, \frac{m_b}{q^2}\, 
\bar s i \sigma_{\mu \nu}q^\nu (1+\gamma_5)  b
+C^{\prime eff}_7(\mu)\, \frac{m_b}{q^2}\, 
\bar s i \sigma_{\mu \nu}q^\nu (1-\gamma_5)
b \, \right) \,\, \bar \ell \gamma^\mu \ell \Bigg\}~\nonumber .
\label{matr}
\end{eqnarray}
\section{The exclusive $B\rightarrow K^* l^+ l^-$ decay} 
Now, our aim is to look at the problem from the hadronic side.
To calculate the decay width, branching ratio, etc., for the exclusive
$B\rightarrow K^* l^+ l^-$ decay, we need the matrix elements
$ \la K^* \vel \bar s \gamma_\mu (1\pm \gamma_5) b \ver B \ra$, and
$\la K^* \vel \bar s i \sigma_{\mu \nu} q^\nu (1\pm\gamma_5) b \ver B \ra$,
Using the parametrization of the form factors as in \cite{R19}, the matrix 
element of the $B\rightarrow K^* l^+ l^-$ decay is obtained as \cite{alsav2}:
\begin{eqnarray}
{\cal M} &=& -\frac{G \alpha_{em}}{2 \sqrt 2 \pi} V_{tb} V_{ts}^*  
\Bigg\{ \bar \ell \gamma^\mu
\ell \left[ 2 A_{tot} \epsilon_{\mu \nu \rho \sigma} \epsilon^{* \nu} 
p_{K^*}^\rho q^\sigma + i
B_{1 \,tot} \epsilon^*_\mu - i B_{2 \,tot} ( \epsilon^* q) 
(p_{B}+p_{K^*})_\mu - 
i B_{3\, tot} (\epsilon^* q)q_\mu \right] \nonumber \\
&+& \bar \ell \gamma^\mu \gamma_5 \ell \left[ 2 C_{tot} \epsilon_{\mu \nu \rho
\sigma}\epsilon^{* \nu} p_{K^*}^\rho q^\sigma + i D_{1\, tot} \epsilon^*_\mu - 
i D_{2\, tot} (\epsilon^* q) (p_{B}+p_{K^*})_\mu - i D_{3\, tot} (\epsilon^* q) 
q_\mu \right] \Bigg\}~,
\label{matr2}
\end{eqnarray}
where $\epsilon^{* \mu}$ is the polarization vector of $K^*$ meson, $p_{B}$ 
and $p_{K^*}$ are four momentum vectors of $B$ and $K^*$ mesons, 
$q=p_B-p_{K^*}$ and

\begin{eqnarray}
A_{tot}&=& A+A' \nonumber \,\, , \\
B_{1\, tot}&=& B_1+B'_1 \nonumber \,\, , \\
B_{2\, tot}&=& B_2+B'_2 \nonumber \,\, , \\
B_{3\, tot}&=& B_3+B'_3 \nonumber \,\, , \\
C_{tot}&=& C+C' \nonumber \,\, , \\
D_{1\, tot}&=& D_1+D'_1 \nonumber \,\, , \\
D_{2\, tot}&=& D_2+D'_2 \nonumber \,\, , \\
D_{3\, tot}&=& D_3+D'_3  \,\, .
\label{hadpar0}
\end{eqnarray}
Here
\begin{eqnarray}
A &=& -C_9^{eff} \frac{V}{m_B + m_{K^*}} - 4 C_7^{eff} \frac{m_b}{q^2} T_1~,
\nonumber\\ 
B_1 &=& -C_9^{eff} (m_B + m_{K^*}) A_1 - 4 C_7^{eff} \frac{m_b}{q^2} (m_B^2 -
m_{K^*}^2) T_2~,  \nonumber \\ 
B_2 &=& -C_9^{eff} \frac{A_2}{m_B + m_{K^*}} - 4 C_7^{eff} \frac{m_b}{q^2} 
\ga T_2 + \frac{q^2}{m_B^2 - m_{K^*}^2} T_3 \dr~,  \nonumber \\
B_3 &=& -C_9^{eff}\frac{ 2 m_{K^*}}{  q^2}(A_3 - A_0) + 4 C_7 
\frac{m_b}{q^2}T_3~,  \nonumber \\ 
C &=& -C_{10} \frac{V}{m_B + m_{K^*}}~,  \nonumber \\  
D_1 &=& -C_{10} (m_B + m_{K^*}) A_1~,  \nonumber \\     
D_2 &=& -C_{10} \frac{A_2}{m_B + m_{K^*}}~,  \nonumber \\ 
D_3 &=& -C_{10} \frac{2 m_{K^*}}{q^2} (A_3 - A_0)~, \nonumber  \\
\label{hadpar1}
\end{eqnarray}
and 
\begin{eqnarray}
A' &=& -C_9^{\prime eff} \frac{V}{m_B + m_{K^*}} - 4 C_7^{\prime eff} \frac{m_b}{q^2} T_1~,
\nonumber\\ 
B'_1 &=& C_9^{\prime eff} (m_B + m_{K^*}) A_1 + 4 C_7^{\prime eff} \frac{m_b}{q^2} (m_B^2 -
m_{K^*}^2) T_2~,  \nonumber \\ 
B'_2 &=& C_9^{\prime eff} \frac{A_2}{m_B + m_{K^*}} + 4 C_7^{\prime eff} \frac{m_b}{q^2} 
\ga T_2 + \frac{q^2}{m_B^2 - m_{K^*}^2} T_3 \dr~,  \nonumber \\
B'_3 &=& C_9^{\prime eff}\frac{ 2 m_{K^*}}{  q^2}(A_3 - A_0) - 4 C_7^{\prime eff} 
\frac{m_b}{q^2}T_3~,  \nonumber \\ 
C' &=& -C'_{10} \frac{V}{m_B + m_{K^*}}~,  \nonumber \\  
D'_1 &=& C'_{10} (m_B + m_{K^*}) A_1~,  \nonumber \\     
D'_2 &=& C'_{10} \frac{A_2}{m_B + m_{K^*}}~,  \nonumber \\ 
D'_3 &=& C'_{10} \frac{2 m_{K^*}}{q^2} (A_3 - A_0)~, \nonumber  \\
\label{hadpar2}
\end{eqnarray}

The hadronic formfactors $V,~A_1,~A_2,~A_0,~T_1,~T_2$ and $T_3$ 
has been calculated in the framework of light-cone QCD sum rules in 
\cite{alsav2,braun}.In our calculations we use the results of \cite{braun} 
where radiative corrections to the leading twist wave function and $SU(3)$
breaking effects are also taken into account. The $q^2$ dependence of the 
form factors can be represented in terms of three parameters as:
\begin{eqnarray}
F(q^2)=\frac{F(0)}{1-a_F \frac{q^2}{m_B^2}+b_F (\frac{q^2}{m_B^2})^2}\, ,
\label{formfac}
\end{eqnarray}
where the values of parameters $F(0)$,$a_F$ and $b_F$ are listed in Table 2.
\begin{table}[h]
    \begin{center}
    \begin{tabular}{|c|c|c|c|}
    \hline
    \hline \hline
                &$F(0)$              &       $a_F$ &  $b_F$\\
    \hline \hline    
    $A_1$       &$0.34\pm 0.05$      &       $0.60$&  $-0.023$ \\
    $A_2$       &$0.28\pm 0.04$      &       $1.18$&  $ 0.281$ \\
    $V  $       &$0.46\pm 0.07$      &       $1.55$&  $ 0.575$ \\
    $T_1$       &$0.19\pm 0.03$      &       $1.59$&  $ 0.615$ \\
    $T_2$       &$0.19\pm 0.03$      &       $0.49$&  $-0.241$ \\
    $T_3$       &$0.13\pm 0.02$      &       $1.20$&  $ 0.098$ \\
    \hline
        \end{tabular}
        \end{center}
\caption{The values of parameters existing in eq.(\ref{formfac}) for 
the various form factors of the transition $B\rightarrow K^*$.} 
\label{Table2}
\end{table}

Light cone QCD sum rules is applicable in the region, 
$m_b^2-q^2\sim$ few $GeV^2$ and this leads to $q^2_{max}\sim 20 \ GeV^2$.
To extend the results to the full physical region, we use the above 
parametrization in such a way that it correctly reproduces light cone 
QCD sum rules results up to $q^2_{max}\sim 20 \ GeV^2$. 
All the errors come from uncertainity of b-quark mass, higher twist 
wave functions, and  variation of Borel paramater are added quadratically. 

Using eq.(\ref{matr2}) we get the double differential decay rate:  
\begin{eqnarray}
\frac{d \Gamma}{d q^2 dz} &=& \frac{G^2 \alpha_{em}^2
\vel V_{tb} V_{ts}^* \ver^2\lambda^{1/2} }{2^{12}
\pi^5 m_B} \Bigg\{ 2 \lambda m_B^4 \Bigg[ 
m_B^2 s ( 1+ z^2) \ga \vel A_{tot} \ver ^2 +\vel C_{tot} \ver ^2 \dr
\Bigg] \nonumber \\
&+& \frac{\lambda m_B^4}{2 r} \Bigg[ \lambda m_B^2 (1-z^2) \ga \vel 
B_{2\, tot} \ver ^2 + \vel D_{2\, tot} \ver ^2 \dr \Bigg]  \nonumber \\
&+& \frac{1}{2 r} \Bigg[ m_B^2 \left\{ \lambda (1- z^2) + 8 r s\right\}
\ga \vel B_{1\, tot} \ver^2 + \vel D_{1\, tot} \ver^2 \dr  
\nonumber \\
&-& 2 \lambda m_B^4
(1-r-s)(1-z^2)  \left\{ Re\ga  B_{1\, tot} B_{2\, tot}^*  \dr + 
Re \ga D_{1\, tot} D_{2\, tot}^*  \dr \right\} 
\Bigg]  \nonumber \\
&-& 8 m_B^4 s\lambda^{1/2} z \Bigg[ 
\left\{ Re\ga  B_{1\, tot} C_{tot}^* \dr + Re\ga A_{tot} D_{1\, tot}^*\dr 
\right\} \Bigg] \Bigg\}~,
\label{dddr}
\end{eqnarray}
where $z=cos \theta$\,, $\theta$ is the angle between the momentum of $\ell$ 
lepton and that of $B$ meson in the center of mass frame of the lepton 
pair, $\lambda = 1+r^2+s^2 -2 r - 2 s - 2 r s$, $r =
\frac{\ds{m_{K^*}^2}}{\ds{m_B^2}}$ and 
$s=\frac{\ds{q^2}}{\ds{m_B^2}}$.

As we noted, $A_{FB}$ and $P_L$ can give more precise information about the 
Wilson coefficients $C_7^{eff},~C_9^{eff}$ and $C_{10}$. They are defined as:
\begin{eqnarray}
A_{FB} (q^2) = \frac{\displaystyle{\int_0^1 dz \frac{d \Gamma}{dq^2 dz} - \int_{-1}^0dz
\frac{d \Gamma}{dq^2 dz}}}{\displaystyle{\int_0^1 dz \frac{d \Gamma}{dq^2 dz}+
\int_{-1}^0dz\frac{d \Gamma}{dq^2 dz}}}
\label{AFB}
\end{eqnarray}
and 
\begin{eqnarray}
P_{L} (q^2) = \frac{\displaystyle{\frac{d \Gamma}{dq^2}(\xi=-1) - 
\frac{d \Gamma}{dq^2}(\xi=-1)}}{\displaystyle{\frac{d \Gamma}{dq^2}(\xi=-1)+
\frac{d \Gamma}{dq^2}(\xi=-1)}}
\label{PL}
\end{eqnarray}
where $\xi= -1 (+1)$ corresponds to the left (right) handed lepton in the
final state. 
After the standard calculation, we get 
\begin{eqnarray}
P_L &=& \frac{1}{\Delta} \Bigg\{ \frac{32}{3} s \lambda m_B^6  Re \ga 
A_{tot} C_{tot}^* \dr +
 \frac{4}{3} \frac{\lambda^2 m_B^6}{r} Re \ga B_{2\, tot} D_{2\, tot}^* \dr +
4 \frac{m_B^2}{r}\left[ \frac{\lambda}{3} + 4 r s
\right] Re \ga B_{1\, tot} D_{1\, tot}^* \dr  \nonumber \\
&-& \frac{4}{3} \frac{\lambda m_B^4}{r} \ga 1 - r -s  \dr  
\left[ Re \ga B_{2\, tot} D_{1\, tot}^* \dr + 
Re \ga B_{1\, tot} D_{2\, tot}^* \dr \right] \Bigg\} \,\, .
\label{PL2}
\end{eqnarray}

The denominator $\Delta$ in eq. (\ref{PL2}) can be obtained from 
eq.(\ref{dddr}) by integration over $z$ of the terms within the curly bracket. 
In our numerical analysis we used the input values given in 
Table~(\ref{input}).
\begin{table}[h]
        \begin{center}
        \begin{tabular}{|l|l|}
        \hline
        \multicolumn{1}{|c|}{Parameter} & 
                \multicolumn{1}{|c|}{Value}     \\
        \hline \hline
        $m_c$                   & $1.4$ (GeV) \\
        $m_b$                   & $4.8$ (GeV) \\
        $\alpha_{em}^{-1}$      & 129           \\
        $\lambda_t$            & 0.04 \\
        $\Gamma_{tot}(B_d)$    & $3.96 \cdot 10^{-13}$ (GeV)   \\
        $m_{B_d}$             & $5.28$ (GeV) \\
        $m_{t}$             & $175$ (GeV) \\
        $m_{W}$             & $80.26$ (GeV) \\
        $m_{Z}$             & $91.19$ (GeV) \\
        $\Lambda_{QCD}$             & $0.214$ (GeV) \\
        $\alpha_{s}(m_Z)$             & $0.117$  \\
        $sin\theta_W$             & $0.2325$  \\
        \hline
        \end{tabular}
        \end{center}
\caption{The values of the input parameters used in the numerical
          calculations.}
\label{input}
\end{table}
\section{Discussion}
Before we present our numerical results, we would like to discuss briefly 
the free parameters of the model III.
This model induces many free parameters, such as $\xi_{ij}^{U,D}$ where 
i,j are flavor indices and it is necessary to restrict them using the 
experimental measurements. 
The contributions of the neutral Higgs bosons $h_0$ and $A_0$ to the Wilson 
coefficient $C_7$ (see the appendix of \cite{alil2} for details) are   
\begin{eqnarray}
C_7^{h_0}(m_W)&=& (V_{tb} V^{*}_{ts} )^{-1}\sum_{i=d,s,b} \bar{\xi}^{D}_{N,bi} 
\,\,\bar{\xi}^{D}_{N,is}\,  \frac{Q_i}{8\, m_i\, m_b}
\nonumber \,\,, \\ 
C_7^{A_0}(m_W)&=& (V_{tb} V^{*}_{ts} )^{-1}\sum_{i=d,s,b} \bar{\xi}^{D}_{N,bi} 
\,\, \bar{\xi}^{D}_{N,is}\, \frac{Q_i}{8\, m_i\, m_b}
\,\, ,
\label{c7A0h0}
\end{eqnarray}
where $m_i$ and $Q_i$ are the masses and charges of the down quarks 
($i=d,\,s,\,b$) respectively. 
These expressions show that the the neutral Higgs bosons can give a large 
contribution to the coefficient $C_7 $ which is in contradiction with 
the CLEO data announced recently  
\cite{cleo2}, 
\begin{eqnarray}
Br (B\rightarrow X_s\gamma)= (3.15\pm 0.35\pm 0.32)\, 10^{-4} \,\, .
\label{br2}
\end{eqnarray}
Such dangerous terms are removed with the assumption that the 
couplings $\bar{\xi}^{D}_{N,is}$($i=d,s,b)$ and $\bar{\xi}^{D}_{N,db}$ are 
negligible to be able to reach the conditions 
$\bar{\xi}^{D}_{N,bb} \,\bar{\xi}^{D}_{N,is} <<1$ and 
$\bar{\xi}^{D}_{N,db} \,\bar{\xi}^{D}_{N,ds} <<1$.
With the discussion given above and using the constraints \cite{alil}, 
coming from the $\Delta F=2$ mixing, the $\rho$ parameter \cite{atwood}, 
and the measurement by CLEO Collaboration we get following restrictions:   
$\bar{\xi}_{N tc} << \bar{\xi}^{U}_{N tt},
\,\,\bar{\xi}^{D}_{N bb}$ and $\bar{\xi}^{D}_{N ib} \sim 0\, , 
\bar{\xi}^{D}_{N ij}\sim 0$, where the indices $i,j$ denote d and s quarks . 
Under this assumption, the Wilson coefficients
$C^{\prime}_{7}$, $C^{\prime}_{9}$ and  $C^{\prime}_{10}$ can be neglected 
compared to unprimed ones and the neutral Higgs 
contributions are supressed because the Yukawa vertices are 
the combinations of $\bar{\xi}^{D}_{N sb}$ and $\bar{\xi}^{D}_{N ss}$.
After these preliminary remarks, let us start with our numerical analysis.

\begin{figure}[htb]
\vskip -1.5truein
\centering
\epsfxsize=3.8in
\leavevmode\epsffile{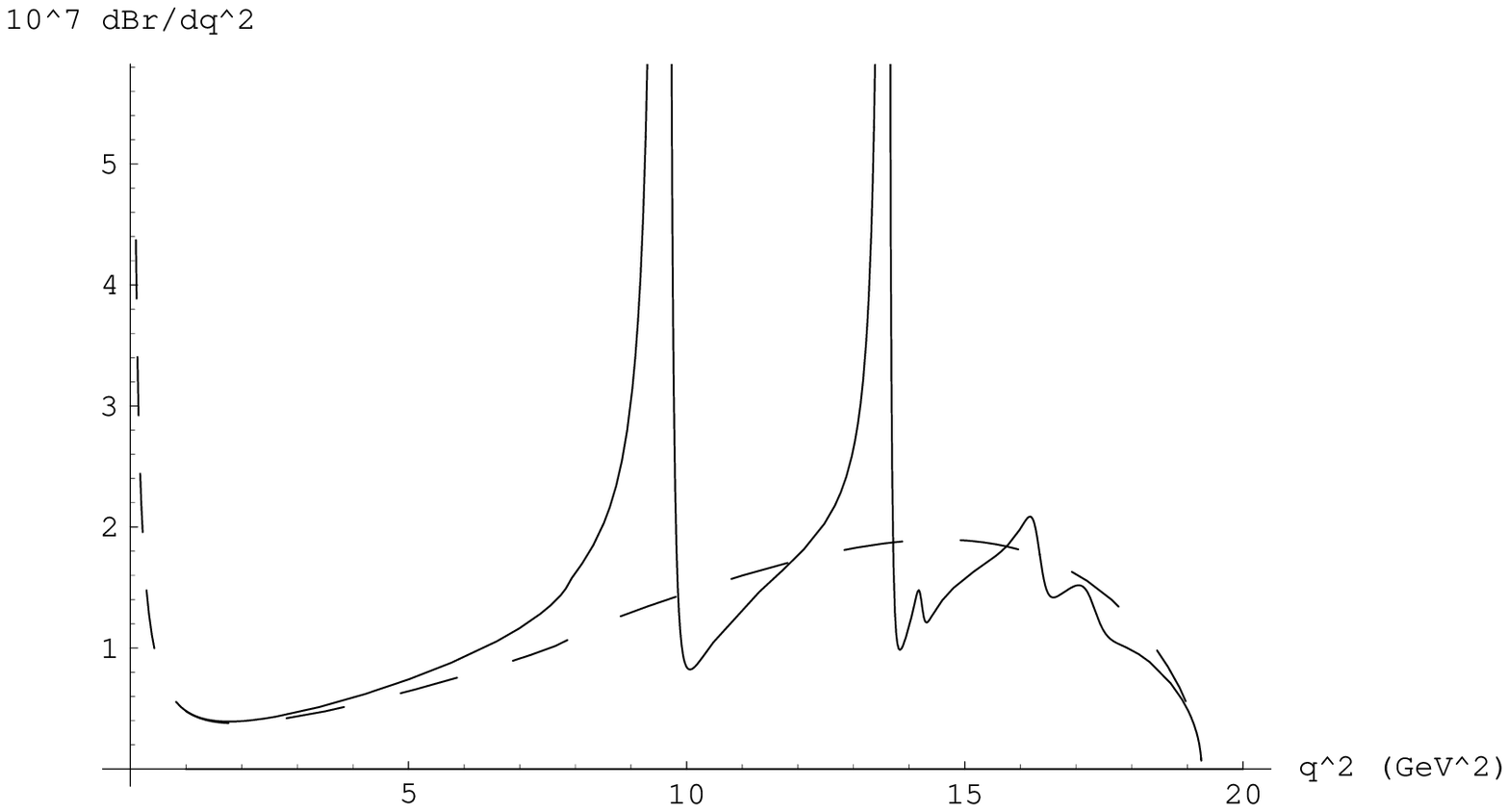}
\vskip -1.5truein
\caption[]{Differential $Br$ as a function of  $q^2$ 
for fixed $\bar{\xi}_{N,bb}^{D}=40\, m_b$ in the region $|r_{tb}|<<1$,
at the scale $\mu=m_b$.
Here solid line corresponds to the model III with LD effects,
and dashed line to the SM withouth LD effects.} 
\label{brbb403q2a}
\end{figure}

\begin{figure}[htb]
\vskip -1.5truein
\centering
\epsfxsize=3.8in
\leavevmode\epsffile{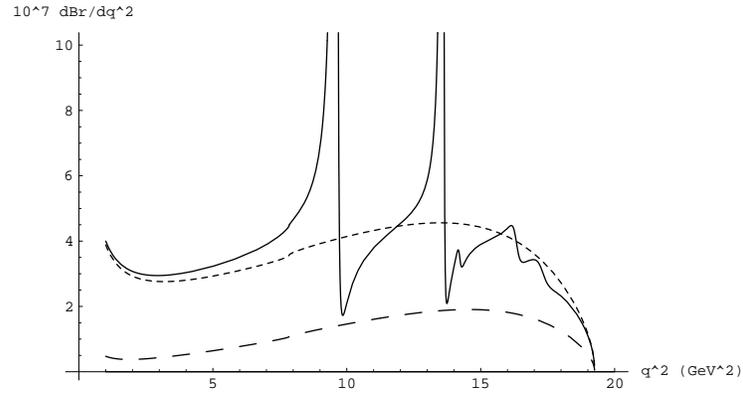}
\vskip -1.5truein
\caption[]{The same as Fig 1, but at the region $r_{tb} >> 1$. Dotted dashed
line corresponds to the model III withouth LD effects.}
\label{brbb403q2b}
\end{figure}

In this section, we study the $q^2$ dependencies of the differential
$Br$, $A_{FB}$ and $P_L$ of the decay $B\rightarrow K^* l^+ l^-$ 
for the selected parameters of the model III
($\bar{\xi}^{U}_{N tt}$,  $\bar{\xi}^{D}_{N bb}$).
In  figs.~\ref{brbb403q2a} and ~\ref{brbb403q2b} we plot the differential 
$Br$ of the decay $B\rightarrow K^* l^+ l^-$ with respect to the dilepton
mass $q^2$ for $\bar{\xi}_{N,bb}^{D}=40\, m_b$ 
and charged Higgs mass $m_{H^{\pm}}=400\, GeV$
at the scale $\mu=m_b$. Fig.~\ref{brbb403q2a} represents the case where
the ratio $|r_{tb}|=|\frac{\bar{\xi}_{N,tt}^{U}}{\bar{\xi}_{N,bb}^{D}}| <<1.$
In this case, it is shown that the differential $Br$ obtained in the 
model III almost coincides with the one calculated in the SM. 
In Fig.~\ref{brbb403q2b}, we present the same dependence as in 
Fig.~\ref{brbb403q2a} for the case $r_{tb}>> 1$ and 
$\bar{\xi}_{N,bb}^{D}=40\, m_b$. Here  
$\bar{\xi}_{N,tt}^{U}$ lies in the restriction region coming from the 
constraints mentioned above. The strong  enhancement is observed compared to 
the SM result especially for the small values of $q^2$. Further, the 
differential $Br$ increases with the increasing $\bar{\xi}_{N,bb}^{D}$ 
for $r_{tb}>> 1$.

\begin{figure}[htb]
\vskip -1.5truein
\centering
\epsfxsize=3.8in
\leavevmode\epsffile{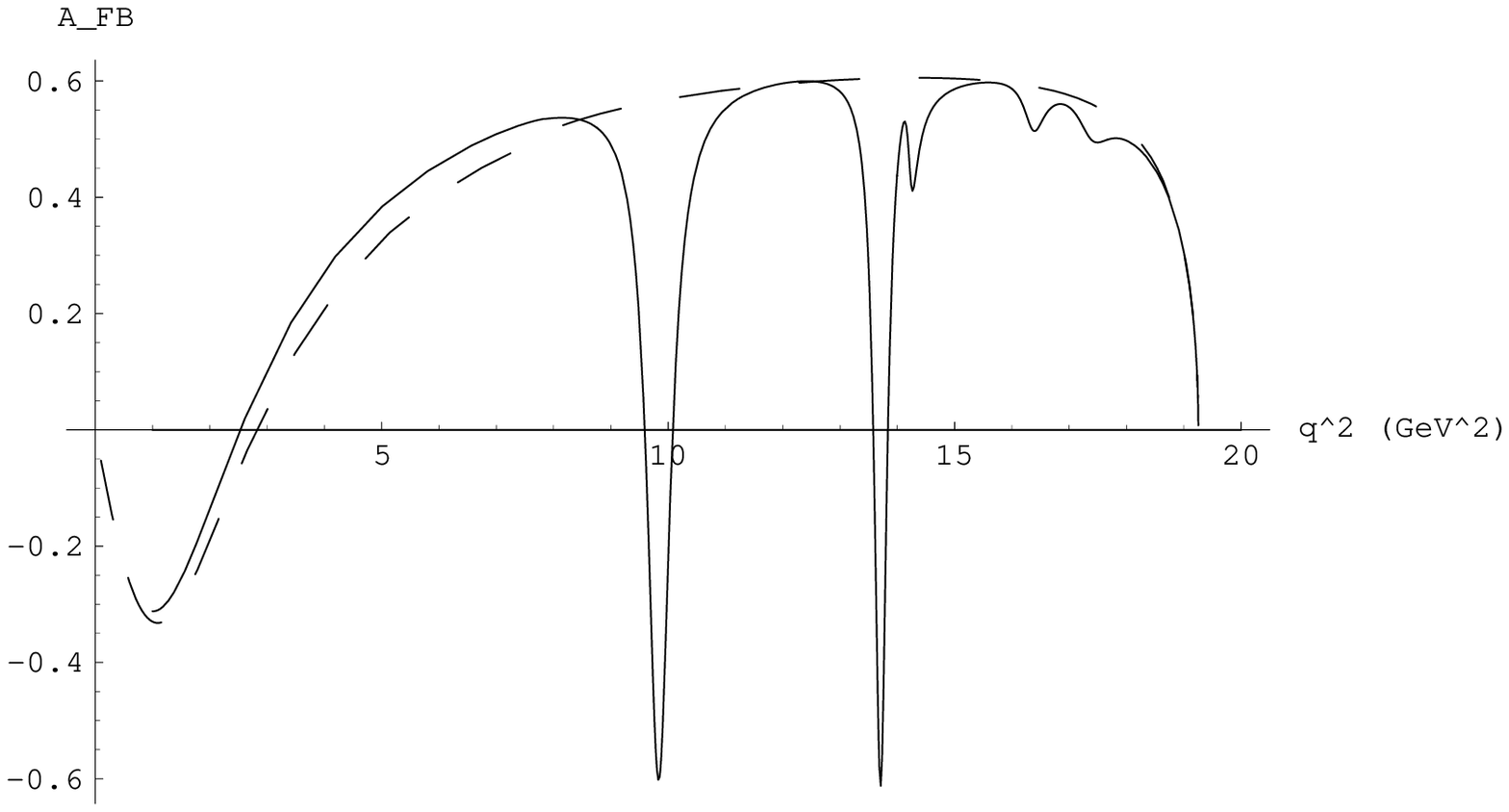}
\vskip -1.5truein
\caption[]{The same as Fig 1, but $A_{FB}$ versus $q^2$ .}
\label{AFBbb403q2a}
\end{figure}

\begin{figure}[htb]
\vskip -1.5truein
\centering
\epsfxsize=3.8in
\leavevmode\epsffile{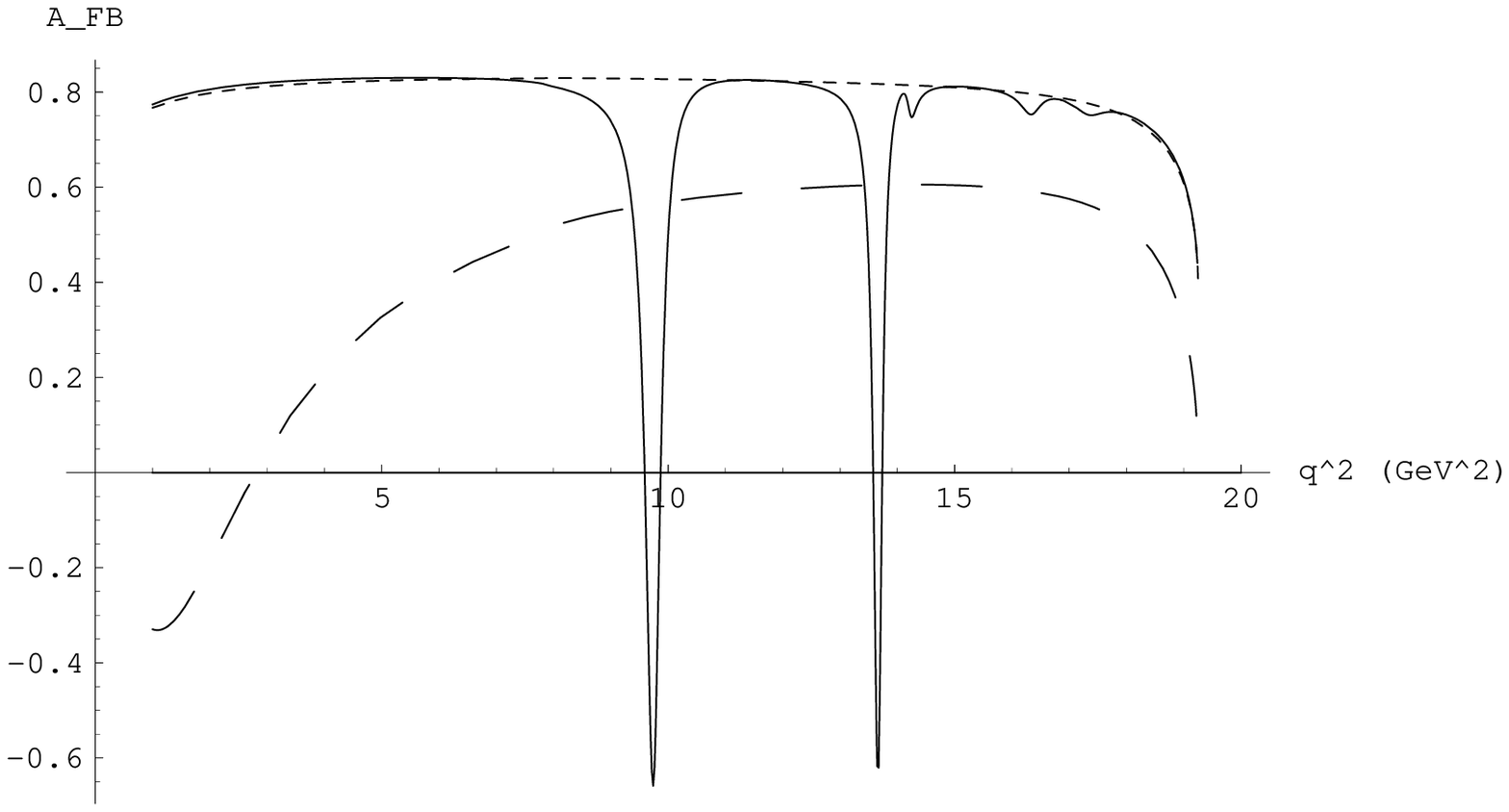}
\vskip -1.5truein
\caption[]{The same as Fig 2, but $A_{FB}$ versus $q^2$ .}
\label{AFBbb403q2b}
\end{figure}

Figs.~\ref{AFBbb403q2a} and  \ref{AFBbb403q2b} show the $q^2$ dependence 
of $A_{FB}$ for $\bar{\xi}_{N,bb}^{D}=40\, m_b$ 
and charged Higgs mass $m_{H^{\pm}}=400\, GeV$ at the scale $\mu=m_b$. 
For $|r_{tb}|<< 1$ (Fig.~\ref{AFBbb403q2a}) $A_{FB}$ changes its sign, 
however for $r_{tb}>> 1$ (Fig.~\ref{AFBbb403q2b}) it is positive withouth LD 
effects. Therefore the determination of the sign of $A_{FB}$ in the region 
$0\le q^2 \le 2.8\, (GeV)^2$ (here the upper limit corresponds to the value 
where $A_{FB}$ change sign in the SM) can give a unique information about
the existence of the model III. Further, $A_{FB}$ becomes more positive with 
increasing $\bar{\xi}_{N,bb}^{D}$. It is interesting to note that the final 
lepton tends to move almost in the direction of $B$ meson with increasing 
$\bar{\xi}_{N,bb}^{D}$ for $r_{tb}>> 1$, especially in the $q^2$ region far 
from LD effects. 

\begin{figure}[htb]
\vskip -1.5truein
\centering
\epsfxsize=3.8in
\leavevmode\epsffile{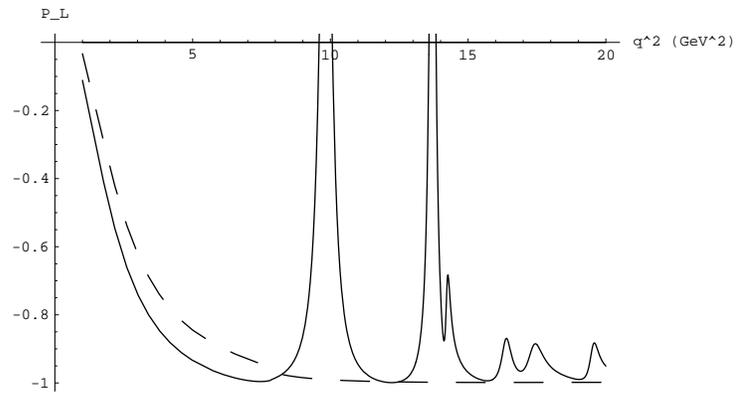}
\vskip -1.5truein
\caption[]{The same as Fig 1, but $P_L$ versus $q^2$ .}
\label{PLbb403q2a}
\end{figure}

\begin{figure}[htb]
\vskip -1.5truein
\centering
\epsfxsize=3.8in
\leavevmode\epsffile{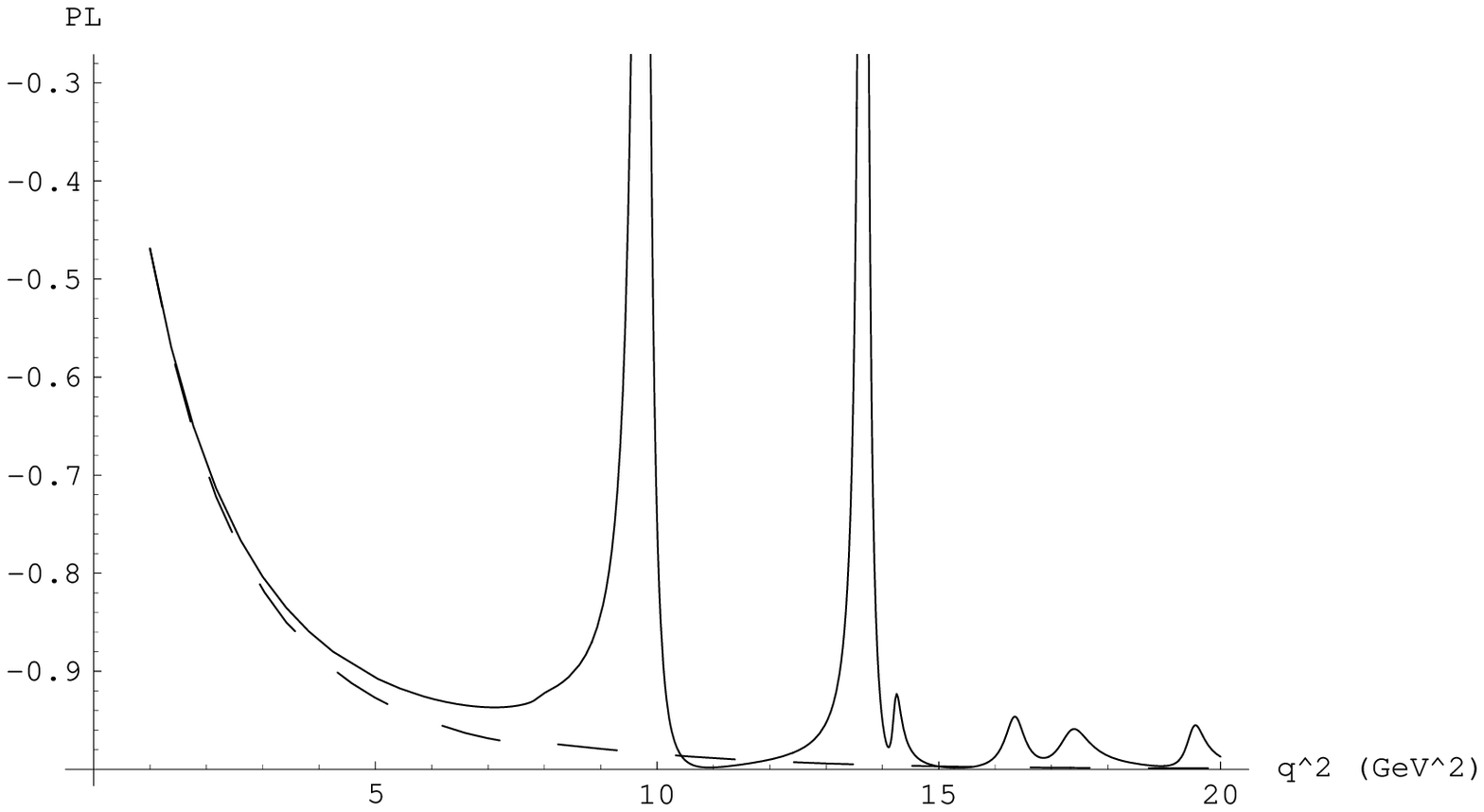}
\vskip -1.5truein
\caption[]{$P_L$ as a function of  $q^2$ 
for fixed $\bar{\xi}_{N,bb}^{D}=40\, m_b$ in the region $r_{tb}>>1$,
at the scale $\mu=m_b$.
Here solid line corresponds to the model III with LD effects,
and dashed line to the model III withouth LD effects.}
\label{PLbb403q2b}
\end{figure}

The behaviour of $P_L$ versus $q^2$ is presented in series of figures 
(\ref{PLbb403q2a} and \ref{PLbb403q2b})
for $\bar{\xi}_{N,bb}^{D}$ 
and charged Higgs mass $m_{H^{\pm}}=400\, GeV$ at the scale $\mu=m_b$. 
$P_L$ becomes more negative in the region $r_{tb}>> 1$ 
(Fig ~\ref{PLbb403q2b}) compared the one in the region $|r_{tb}|<< 1$ 
(Fig ~\ref{PLbb403q2a}). This tendency increases with increasing  
$\bar{\xi}_{N,bb}^{D}$.

Now we would like to present the values of $Br$ for the 
$B \rar K^* l^+ l^-$ decay in the SM and model III, without the LD effects.
After integrating over $q^2$, we get
\begin{eqnarray}
Br(B \rar K^* l^+ l^-)=0.278\pm 0.050 \times 10^{-5} \,\,\, (SM)
\label{Br2SM}
\end{eqnarray}
and for the model III
\begin{eqnarray}
Br(B \rar K^* l^+ l^-) = \left\{ \begin{array}{ll}
~ 0.278\pm 0.050 \times 10^{-5}   & (|r_{tb}|<<1) \\ \\
~ 0.990\pm 0.050 \times 10^{-5}   & (r_{tb}>>1 \,,\,\bar{\xi}_{N,bb}^{D}=40\, m_b ).
\end{array} \right.
\label{Br2}
\end{eqnarray}
The $Br$ in the SM is almost equal to the model III value for 
$|r_{tb}|<<1$. However, the strong enhancement can be observed for 
$r_{tb}>>1$, especially with increasing $\bar{\xi}_{N,bb}^{D}$.  
Using the CLEO data for the inclusive branching ratio 
$Br(B \rar X_s l^+ l^-) \leq 0.42 \,\,10^{-4}$ \cite{glenn}, it is possible to
estimate an upper limit for $\bar{\xi}_{N,bb}^{D}$, which is 
$\bar{\xi}_{N,bb}^{D} \leq 40\, m_b$. 

In conclusion, we analyse the selected 
model III parameters ( $\bar{\xi}_{N,bb}^{D}$, $\bar{\xi}_{N,tt}^{U}$ ) 
dependencies of the differential $Br$ ,$A_{FB}$ and $P_L$ of the decay
$B\rightarrow K^* l^+ l^-$. 
We obtain that the strong enhancement of the differential $Br$ is possible 
in the framework of the model III and observe that $A_{FB}$ and $P_L$ 
are very sensitive to the model III parameters 
($\bar{\xi}_{N,bb}^{D}$,  $\bar{\xi}_{N,tt}^{U}$). 
Therefore, their experimental investigations ensure a crucial test
for new physics.

\newpage

\newpage
\begin{thebibliography}{1}
\bibitem{Hewett} J. L. Hewett, in Proc. of the $21^{st}$ Annual SLAC Summer 
Institute, ed. L. De Porcel and C. Dunwoode, SLAC-PUB-6521 (1994)
%
\bibitem{cleo} M. S. Alam {\it et. al.}
{\bf CLEO} Collaboration, {\it Phys. Rev.Lett.} {\bf 74} (1995) 2885; 
%
\bibitem{rammar} R. Ammar {\it et. al.}
{\bf CLEO} Collaboration, {\it Phys. Rev. Lett.} {\bf 71} (1993) 674.
%
\bibitem{R4} W. -S. Hou, R. S. Willey and A. Soni,
{\it Phys. Rev. Lett.} {\bf 58} (1987) 1608.

\bibitem{R5} N. G. Deshpande and J. Trampetic,
{\it Phys. Rev. Lett.} {\bf 60} (1988) 2583.

\bibitem{R6} C. S. Lim, T. Morozumi and A. I. Sanda,
{\it Phys. Lett.} {\bf B218} (1989) 343.

\bibitem{Grinstein1} B. Grinstein, M. J. Savage and M. B. Wise,
{\it Nucl. Phys.} {\bf B319} (1989) 271.

\bibitem{R8} C. Dominguez, N. Paver and Riazuddin, 
{\it Phys. Lett.} {\bf B214} (1988) 459.

\bibitem{R9} N. G. Deshpande, J. Trampetic and K. Ponose,
{\it Phys. Rev.} {\bf D39} (1989) 1461.

\bibitem{donnel} P. J. O'Donnell and H. K. Tung,
{\it Phys. Rev.} {\bf D43} (1991) 2067.

\bibitem{R12} N. Paver and Riazuddin,
{\it Phys. Rev.} {\bf D45} (1992) 978.

\bibitem{ali} A. Ali, T. Mannel and T. Morozumi,
{\it Phys. Lett.} {\bf B273} (1991) 505.

\bibitem{R14} A. Ali, G. F. Giudice and T. Mannel,
{\it Z. Phys.} {\bf C67} (1995) 417.

\bibitem{R15} C. Greub, A. Ioannissian and D. Wyler,
{\it Phys. Lett.} {\bf B346} (1995) 145; \\
D. Liu {\it Phys. Lett.} {\bf B346} (1995) 355; \\
G. Burdman, {\it Phys. Rev.} {\bf D52} (1995) 6400: \\
Y. Okada, Y. Shimizu and M. Tanaka {\it Phys. Lett.} {\bf B405} (1997) 297.

\bibitem{buras} A. J. Buras and M. M\"{u}nz,
{\it Phys. Rev.} {\bf D52} (1995) 186.

\bibitem{R17} N. G. Deshpande, X. -G. He and J. Trampetic,
{\it Phys. Lett.} {\bf B367} (1996) 362.
%
\bibitem{R10} W. Jaus and D. Wyler,
{\it Phys. Rev.} {\bf D41} (1990) 3405.
%
\bibitem{R18} R. Casalbuoni, A. Deandra, N. Di Bartolemo, 
R. Gatto and G. Nardulli,\\
{\it Phys. Lett.} {\bf B312} (1993) 315.

\bibitem{R19} P. Colangelo, F. De Fazio, P. Santorelli and E. Scrimieri,
{\it Phys. Rev.} {\bf D53} (1996) 3672.

\bibitem{R21} W. Roberts,
{\it Phys. Rev.} {\bf D54} (1996) 863.

\bibitem{alsav2} T. M. Aliev, A. \"{O}zpineci and M.Savc{\i},
{\it Phys. Rev.} {\bf D56} (1997) 4260.
%
\bibitem{braun} P. Ball and V. Braun, 
{\it Phys. Rev.} {\bf D57} (1998) 4260.

\bibitem{soni} D. Atwood, L. Reina and A. Soni, 
{\it Phys. Rev.} {\bf D53} (1996) 119.
%
\bibitem{ciuchini2} M. Ciuchini, G. Degrassi, P. Gambino and G. I. Giudice,
{\it Nucl. Phys.} {\bf B527} (1998) 21.
%
\bibitem{gudalil} T. M. Aliev, G. Hiller and E. O. Iltan,
{\it Nucl. Phys.} {\bf B515} (1998) 321.
%
\bibitem{Grinstein2} B. Grinstein, R. Springer, and M. Wise,
{\it Nucl. Phys.} B{\bf 339} (1990) 269; R. Grigjanis, P.J. O'Donnel,
M. Sutherland and H. Navelet, {\it Phys. Lett.} B{\bf 213} (1988) 355; 
{\it Phys. Lett.} B{\bf 286} (1992) E, 413; 
G. Cella, G. Curci, G. Ricciardi and 
A. Vicer\'e, {\it Phys. Lett.} B{\bf 325} (1994) 227, 
{\it Nucl. Phys.} B{\bf 431} (1994) 417. 
%
\bibitem{misiak}
M. Misiak, {\it Nucl. Phys.} B{\bf 393} (1993) 23, Erratum B{\bf 439} (1995) 
461.
%
\bibitem{alil} T. M. Aliev, and E. Iltan, hep-ph/9803272, (1998).
%
\bibitem{cho} P. Cho and Misiak, {\it Phys. Rev.} D {\bf 49} (1994) 5894.
%
\bibitem{greub2} F. M. Borzumati and C. Greub, 
{\it Phys. Rev.} {\bf D58} (1998) 0784004
%
\bibitem{int1} G. Buchalla, G. Isidori and S. J. Rey,
{\it Nucl. Phys.} {\bf B511} (1998) 594.
%
\bibitem{int3} Z. Ligeti, I. W. Steward and M. B. Wise ,
{\it Phys. Lett.} {\bf B420} (1998) 359.
%
\bibitem{int4} F. Kr\"{u}ger and L. M. Sehgal 
{\it Phys. Lett.} {\bf B380} (1996) 199.
%
\bibitem{kagan} A. Kagan, M. Neubert, hep-ph/9805303, 1998.
%
\bibitem{cabbibo}
N. Cabbibo and R. Gatto, {\it Phys. Rev.} {\bf D124} (1961) 1577\, , \\
H. Burkhardt, F.Jegerlehner, G. Penso and C. Verzegnassi,  
{\it Z. Phys.} {\bf C43} (1989) 497.
%
\bibitem{sehgal2} F. Kr\"{u}ger and L. M. Sehgal 
{\it Phys. Lett.} {\bf B380} (1996) 199.
%
\bibitem{alil2} T. M. Aliev, and E. Iltan, 
{\it Phys. Rev.} {\bf D58} (1998) 095014.
%
\bibitem{cleo2} M. S. Alam, CLEO Collaboration, to appear in ICHEP98 
Conference (1998)
%
\bibitem{atwood} D. Atwood, L. Reina and A. Soni, 
{\it Phys. Rev.} {\bf D55} (1997) 3156.
%
\bibitem{glenn} Glenn et al., {\it Phys. Rev.} D {\bf 80} (1998) 2289.
%
\end{thebibliography}
\end{document}